\documentclass[prb,reprint,aps,amsmath,amssymb,showpacs, superscriptaddress]{revtex4-2}
\usepackage{graphicx}

\begin{document}
\preprint{}
\draft

\title{Interference between distinguishable photons}

\author{Manman Wang}\thanks{These authors contributed equally to this work}
\affiliation{Beijing Academy of Quantum Information Sciences, Beijing 100193, China.}
\affiliation{Institute of Physics, Chinese Academy of Sciences, Beijing 100190, China.}
\affiliation{University of Chinese Academy of Sciences, Beijing 101408, China.}

\author{Yanfeng Li}\thanks{These authors contributed equally to this work}
\affiliation{Beijing Academy of Quantum Information Sciences, Beijing 100193, China.}
\affiliation{Institute of Physics, Chinese Academy of Sciences, Beijing 100190, China.}
\affiliation{University of Chinese Academy of Sciences, Beijing 101408, China.}

\author{Hanqing Liu}
\affiliation{Institute of Semiconductors, Chinese Academy of Sciences, Beijing 100083, China.}
\affiliation{College of Materials Science and Opto-Electronic Technology, University of Chinese Academy of Sciences, Beijing 101408, China.}

\author{Haiqiao Ni}
\affiliation{Institute of Semiconductors, Chinese Academy of Sciences, Beijing 100083, China.}
\affiliation{College of Materials Science and Opto-Electronic Technology, University of Chinese Academy of Sciences, Beijing 101408, China.}

\author{Zhichuan Niu}
\affiliation{Institute of Semiconductors, Chinese Academy of Sciences, Beijing 100083, China.}
\affiliation{College of Materials Science and Opto-Electronic Technology, University of Chinese Academy of Sciences, Beijing 101408, China.}

\author{Chengyong Hu} \email[Corresponding author: ]{cyhu03@gmail.com}
\affiliation{Beijing Academy of Quantum Information Sciences, Beijing 100193, China.}

\begin{abstract}
\noindent Two-photon interference (TPI) lies at the heart of photonic quantum technologies. TPI
is generally regarded as quantum interference stemming from the indistinguishability of identical photons,
hence a common intuition prevails that TPI would disappear if photons are distinguishable.
Here we disprove this perspective and uncover the essence of TPI.
We report the first demonstration of TPI between distinguishable photons
with their frequency separation up to $10^4$ times larger than their linewidths.
We perform time-resolved TPI between an independent laser and single photons with
ultralong coherence time ($>10\ \mu$s). We observe a maximum TPI visibility of $72\%\pm 2\%$
well above the $50\%$ classical limit indicating the quantum feature, and simultaneously
a broad visibility background and a classical beat visibility of less than $50\%$ reflecting the
classical feature. These visibilities are independent of the photon frequency separation and
show no difference between distinguishable and indistinguishable photons.
Based on a general wave superposition model, we derive the cross-correlation functions
which fully reproduce and explain the experiments. Our results reveal that TPI as the fourth-order interference
arises from the second-order interference of two photons within the mutual coherence time
and TPI is not linked to the photon indistinguishability.
This work provides new insights into the nature of TPI with great implications in both quantum optics
and photonic quantum technologies.

\end{abstract}

\date{\today}

\maketitle

\noindent Two-photon interference (TPI) also known as the Hong-Ou-Mandel (HOM) interference \cite{hong87}
is a fundamental phenomenon in quantum optics and finds wide applications in quantum
information science and technology \cite{bouwmeester97, pan98, briegel98, pan12, xu20, pelucchi22, giovannetti11, pirandola18}.
When two identical photons that have the same frequency,
polarization, and temporal and spatial extent are incident on a 50/50 beam splitter
from two input ports, they will always bind together and leave the beam splitter from the same
output port. According to the common views \cite{bouchard21}, TPI stems from the indistinguishability of identical
photons and the photon bunching is regarded as a manifestation of the bosonic nature of light.
This perspective has guided a series of subsequent experiments to make indistinguishable photons
from either independent or dissimilar photon sources \cite{santori02, kaltenbaek06, bennett09, deng19, zhai22, you22} for applications in quantum communications \cite{xu20} and quantum networks \cite{kimble08, wehner18, lu21}.

However, there are a few works reporting TPI also occurs between distinguishable photons and suggesting
photon indistinguishability is not necessary at the beam splitter but at the detectors which erase
the distinguishable information of photons \cite{pittman97, kim99, legero04, zhao14, tamma15, yard24}.
TPI is observed with the photon frequency separation larger than their linewidths
by $8.6$ times \cite{legero04}, $16$ times \cite{zhao14}, and $1.8$ times \cite{yard24}, respectively.
Recently, Jones et al \cite{jones20} demonstrated multiparticle interference of pairwise
distinguishable photons where the photon pairs remain distinguishable after detection with the detectors
resolving only the presence of a photon, not the internal mode structure.
Very recently, Seron et al \cite{seron23} proved theoretically that boson bunching in a
multimode interferometer is not maximized by indistinguishable particles using the
theory of matrix permanent. All these works imply that TPI may not be directly linked to
the indistinguishability of identical photons, challenging the common perspective.
Further investigation is required to clarify the essence of TPI.

Interference is intrinsically a classical wave phenomenon based on the superposition principle in
both classical and quantum physics. This is also true for TPI from the viewpoint of
wave-like feature of photons.
In this work, we uncover the essence of TPI by performing time-resolved
TPI experiments between an independent laser and single photons with ultralong coherence time.
The photons from two different sources are tunable from indistinguishable to spectrally distinguishable
with their frequency separation up to $10^4$ times larger than their linewidths ($<100\ $kHz).
In most TPI experiments reported so far, single photons have short coherence time
limited by the photon correlation time or twice the emitter's radiative lifetime.
As a result, it is hard to distinguish between interference feature and single-photon characteristics.
Here we coherently convert laser light into single photons using a single quantum dot (QD) coupling
to an optical microcavity in the Purcell regime. Such single photons
inherit the incident laser's ultralong coherence ($>10\ \mu$s) which is five orders
of magnitude longer than the photon correlation time, and allow us to discriminate
quantum and classical effects of TPI and related interference beat.
We observe a maximum TPI visibility of $72\%\pm 2\%$
well above the $50\%$ classical limit, and simultaneously
a broad visibility background and a classical beat visibility of less than $50\%$.
These visibilities are independent of the photon frequency separation
and show no difference between indistinguishable and distinguishable photons.
Based on these results and a general wave superposition theory, we identify
that TPI as the fourth-order interference arises from the
second-order interference of two photons arriving on the beam splitter
within the mutual coherence time no matter whether
two photons are distinguishable or indistinguishable.
Correspondingly the photon bunching induced by interference
takes root in path indistinguishability, not photon indistinguishability, disproving
the common perspective on TPI and its relation with photon indistinguishability.

\begin{figure}[ht]
\centering
\includegraphics* [bb= 106 383 520 658, clip, width=8cm, height=5.4cm]{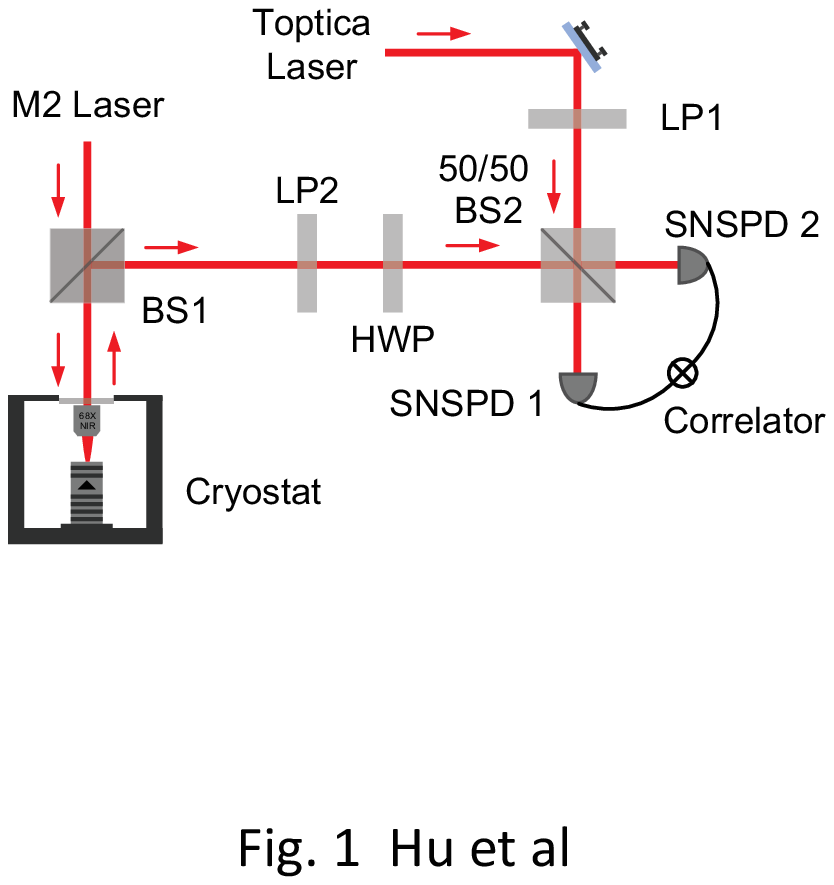}
\caption{(color online) Experimental setup for time-resolved TPI measurement between single photons
and an independent laser (Toptica CTL-950).
Single photons with ultralong coherence time are generated by a single QD in an
optical microcavity driven by a tunable cw Ti:sapphire laser (M SQUARED) with a linewidth $<100\ $kHz.
The sample is placed inside a closed-cycle cryostat (attocube attoDry 800).
Time-resolved correlation measurements are performed with an 8-channel time-correlated single photon counter (Swabian
Instruments)
with a time jitter of $3\ $ps.
BS, $\mathrm{BS_A}$ and $\mathrm{BS_B}$: beam splitters, LP1 and LP2: linear polarizers, HWP: half-wave plate, SNSPD1 and SNSPD2: superconducting nano-wire single-photon detectors (Scontel) with a time jitter of $20\ $ps.}
\label{fig1}
\end{figure}

We designed and fabricated a pillar microcavity containing a single self-assembled In(Ga)As QD
resonantly coupling to the fundamental cavity mode with the cooperativity parameter
$C=2g^2/(\kappa \gamma_{\perp}) \gg 1$ and critical photon number
$n_0=\gamma_{\perp}\gamma_{\parallel}/(4g^2) \ll 1$, where $g$ is the QD-cavity interaction strength,
$\kappa$ is the cavity photon decay rate, $\gamma_{\parallel}$ is the QD spontaneous emission rate into leaky modes,
$\gamma_{\perp}=\gamma_{\parallel}/2+\gamma^*$ is the QD polarization decay rate, and
$\gamma^*$ is the QD pure dephasing rate.
Such design allows the incident laser light interacts with the QD deterministically,
cavity-enhanced coherent scattering and strong nonlinearity at the single-QD and single-photon
level. The cavity is defined by two mirrors made up of 18 and 30 pairs of
GaAs/Al$_{0.9}$Ga$_{0.1}$As distributed Bragg reflectors (DBRs), respectively.
The two DBR mirrors are made asymmetric in the realistic devices such that the leakage rate
from the top mirror can balance the total leakage rates from the bottom mirror,
cavity side and background absorption resided in materials.
This cavity structure mimics a double-sided cavity with zero reflectivity at the center of
the fundamental cavity mode. The details for sample growth and device fabrication can be found
in our recent work \cite{li24}.
The cavity quantum electrodynamics (CQED) parameters for the sample used in this work
are $g/2\pi=4.7\ $GHz, $\kappa/2\pi=36.8\ $GHz, $\gamma_{\parallel}/2\pi = 0.35\ $GHz and
$\gamma^*/2\pi \simeq 0\ $GHz. So the cooperativity parameter is $C=6.9$ and the critical
photon number is $n_0=6.9\times 10^{-4}$.

\begin{figure*}[ht]
\centering
\includegraphics* [bb= 130 478 448 680, clip, width=12cm, height=7.4cm]{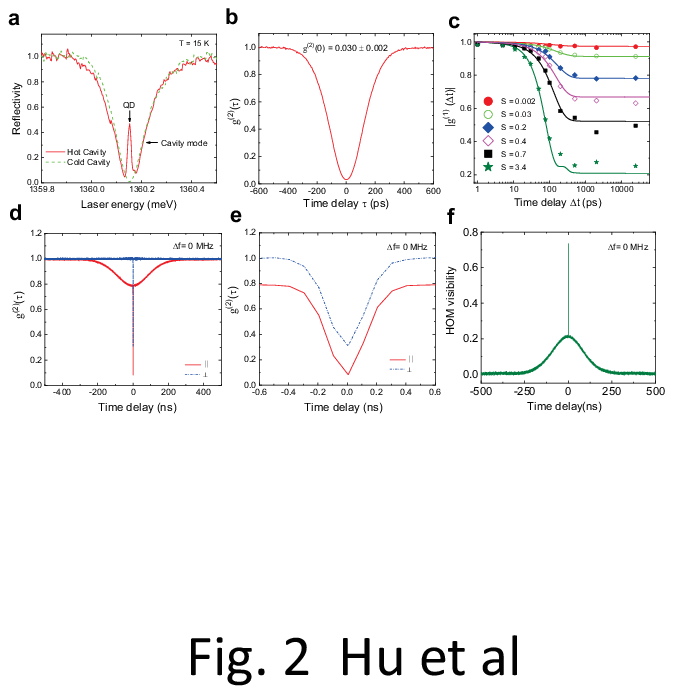}
\caption{(color online) Generation of single photons
with ultralong coherence time ($>10\ \mu$s).
(a) Reflection spectra measured by scanning the laser's wavelength. The QD-cavity coupled system (i.e., hot cavity)
at low driving fields (red, solid line) exhibits different reflection spectra from the cold or empty cavity at
high driving fields where the QD gets saturated (blue, dashed line).
(b) Measured second-order autocorrelation $\mathrm{g}^{(2)}(\tau)$ of
reflected light with the laser frequency fixed at the QD resonance. $\mathrm{g}^{(2)}(0)=0.030\pm 0.002$
is achieved.
(c) Measured degree of first-order coherence $|\mathrm{g}^{(1)}(\tau)|$ versus
time delay at different driving powers with a Michelson interferometer.
The solid curves represent calculated results using the master equation \cite{li24}.
We choose a low driving power with the saturation parameter $S=0.01$ for TPI experiments in this work.
(d) Cross-correlation measurements for the superimposed light of laser and single photons in cross-polarization (blue dot)
and parallel-polarization (red solid) configurations.
(e) Magnified detail of the single-photon anti-bunching dip at zero time delay.
(f) HOM visibility versus the time delay. In (d)-(f), the photon frequency separation is kept
at $\Delta f=0\ $MHz and the intensity ratio of laser to single photons is fixed to $0.2$.}
\label{fig2}
\end{figure*}

The sample was placed inside a closed-cycle cryostat (see Fig. 1) and the QD transition was adjusted in resonance
with the cavity mode by temperature tuning. A cw tunable Ti:saphhire laser (M SQUARED) with the linewidth
$<100\ $kHz was used to drive the cavity. We monitor the intensity or correlations of the reflected light.
Fig. 2(a) presents the coherent reflection spectra measured by scanning the laser's wavelength.
Nearly zero reflectivity ($R=0.89\%$) at the center frequency
of cavity mode is observed at higher driving powers when QD get saturated [Fig. 2(a), blue dashed line] .
The QD transition induces a sharp reflection peak ($R=46.6\%$)
inside the cavity-mode resonance [Fig. 2(a), red solid line] at lower driving powers.

Fixing the laser frequency on the QD resonance, we measure the second-order autocorrelation
function of reflected light and achieve $\mathrm{g}^{(2)}(0)=0.030\pm 0.002$ [see Fig. 2(b)]
at low driving fields with the saturation parameter $S<1$.
The reflected light consists of a superposition of the driving field and the cavity output field.
The driving laser field shows Poissonian statistics while the cavity output field exhibits
super-bunching \cite{li24} due to photon-induced tunneling \cite{faraon08, majumdar12}
and multi-photon scattering \cite{shen07},
so the common picture that a single QD can only scatter (or absorb and emit) single photons
is not the case at low driving fields. We identify that fully destructive interference
between the driving field and the cavity output field erases the two-photon probability
amplitude in reflected light and converts the driving laser light into single photons \cite{li24},
affirming the interference picture on antibunching in resonance fluorescence proposed
40 years ago \cite{dalibard83, masters23}.

At low driving fields with $S<1$, the cavity output intensity is much weaker than the driving field, so
the first-order coherence of reflected single photons is determined by the driving field.
Fig. 2(c) presents the degree of first-order coherence $\mathrm{g}^{(1)}(\tau)$ of reflected light versus
time delay measured with a Michelson interferometer at different driving powers. There are two coherence times observed,
$\tau_{c1} \simeq 115$ ps and  $\tau_{c2}> 24.5$ ns which is limited by the longest path delay of interferometer.
The short 115-ps coherence time which is the photon correlation time twice the QD radiative lifetime $57\ $ps
stems from the incoherent cavity output field due to quantum fluctuations \cite{li24},
while the long coherence time comes from
the driving field with ultra-long coherence time ($>10\ \mu$s).
The incoherent cavity output reduces the degree of first-order coherence by its intensity fraction in reflected light,
however, this reduction is negligible  at low driving fields.
Single photons with subnatural linewidth were also reported
in previous work \cite{nguyen11, matthiesen12, hanschke20, phillips20}.

\begin{figure}[ht]
\centering
\includegraphics* [bb= 135 367 488 538, clip, width=8.5cm, height=4.1cm]{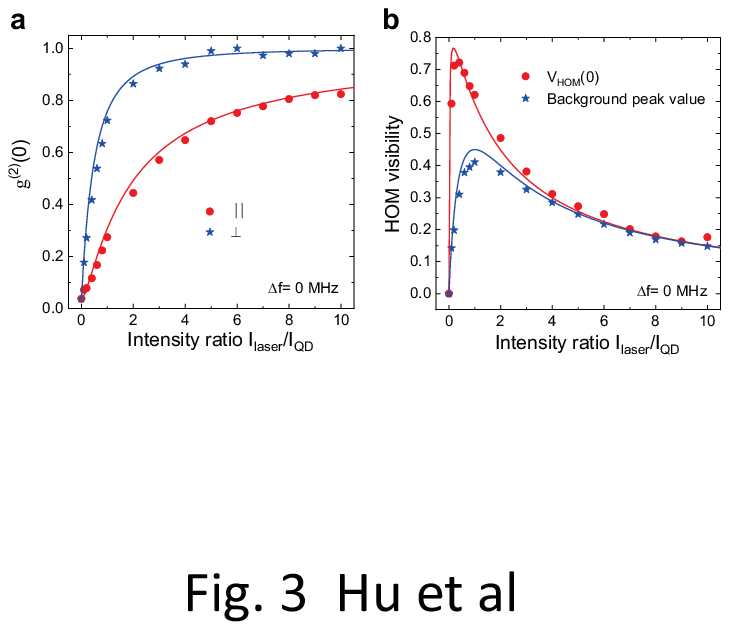}
\caption{(color online) Dependence of cross-correlation measurements on the intensity
ratio of laser light to single photons.
(a) $\mathrm{g}^{(2)}_{\parallel}(0)$ and $\mathrm{g}^{(2)}_{\perp}(0)$ versus the intensity ratio.
(b) $V_{HOM}(0)$ and the peak value of visibility background versus the intensity ratio.
Dots represent the experimental data and solid lines represent the calculated results in (a)
using Eqs.(\ref{eq1})-(\ref{eq2}) and (b) using Eq. (\ref{eq3}).
The photon frequency separation is kept at $\Delta f=0\ $MHz. }
\label{fig3}
\end{figure}

We use another cw tunable laser (Toptica CTL-950) to perform TPI with single photons generated
by the cw tunable Ti:saphhire laser (M SQUARED) (see Fig. 1). The two lasers are independent
of each other. We choose a low driving-field strength with the saturation parameter $S=0.01$
in the Heitler regime \cite{nguyen11, matthiesen12} such that the reflected single photons
have a high degree of first-order coherence with $|\mathrm{g}^{(1)}(\tau)|>95\%$ [see Fig. 2(c)]
for $\tau$ much less than the coherence time.
Fig. 2(d) presents the measured cross-correlation functions $\mathrm{g}^{(2)}_{\perp}(\tau)$
and $\mathrm{g}^{(2)}_{\parallel}(\tau)$ for the superimposed light of laser and single photons in
cross- and parallel-polarization configurations. The photon frequency separation is set to zero,
so the photons from laser and single photons are indistinguishable with each other.

In cross-polarization configuration, no interference is expected and we observe
a sharp dip at $\tau=0$ with a half width of $115\ $ps [see Figs. 2(d) and 2(e), blue curves].
This is the single-photon antibunching dip with a correlation time of $115\ $ps,
twice the QD radiative lifetime ($57\ $ps).

In parallel-polarization configuration, besides the sharp single-photon dip at $\tau=0$, we observe
a broad dip with a half width of $150\ $ns [see Fig. 2(d), red curve], which
is the well-known HOM dip due to TPI. From it, we get the mutual coherence
time $150\ $ns between laser light and single photons.

The single-photon dip is shifted downwards by the HOM dip. The HOM (or TPI) visibility is usually defined as
$V_{HOM}(\tau)=[\mathrm{g}^{(2)}_{\perp}(\tau)-\mathrm{g}^{(2)}_{\parallel}(\tau)]/\mathrm{g}^{(2)}_{\perp}(\tau)$.
From this definition, we get the maximum TPI visibility $V_{HOM}(0)=72\%\pm 2\%$ [Fig. 2(f)]
which is well above the $50\%$ classical limit, indicating the quantum feature of TPI.
The broad visibility background with peak value less than $22\%\pm 1\%$ reflects the classical feature of TPI.

Fig. 3 displays the dependence of $\mathrm{g}^{(2)}_{\perp}(\tau)$, $\mathrm{g}^{(2)}_{\parallel}(\tau)$ and $V_{HOM}(\tau)$
on the intensity ratio $\eta$ of laser to single photons.
In order to understand the above results, we use a general wave superposition model to
work out the cross-correlation functions of the superimposed light of laser light and single
photons [see Eqs.(S5) and (S6) in Supplemental Information] in cross-polarization configuration
\begin{widetext}
\begin{equation}
\mathrm{g}^{(2)}_{\perp}(\tau)=\frac{1}{N}\Biggl \{\eta^2RT\mathrm{g}_L^{(2)}(\tau)+RT\mathrm{g}_{SP}^{(2)}(\tau)
+\eta(R^2+T^2)\Biggl\},
\label{eq1}
\end{equation}
and in parallel-polarization configuration
\begin{equation}
\begin{split}
\mathrm{g}^{(2)}_{\parallel}(\tau)=\frac{1}{N}\Biggl\{\eta^2RT\mathrm{g}_L^{(2)}(\tau)+RT\mathrm{g}_{SP}^{(2)}(\tau)
+\eta(R^2+T^2)-2\eta RTV_0|\mathrm{g}_{SP}^{(1)}(\tau)||\mathrm{g}_L^{(1)}(\tau)| \cos (2\pi\Delta f\tau)\Biggl\},
\end{split}
\label{eq2}
\end{equation}
\end{widetext}
where $N=(1+\eta^2)RT+\eta(R^2+T^2)$ is the normalization factor and
$\eta=I_{laser}/I_{QD}$ is the intensity ratio of laser light to single photons, $R$ and $T$
are the reflection and transmission intensity coefficients of BS2
with nominal values $R=T=50\%$. The parameter $V_0$ is
introduced to consider the mode overlap on BS2 (see Supplemental Information).
$\mathrm{g}_{L}^{(2)}(\tau)=1$ and $\mathrm{g}_{SP}^{(2)}(\tau)$ are the second-order auto-correlation functions
of laser light and single photons, respectively. $\mathrm{g}_{L}^{(1)}(\tau)$ and $\mathrm{g}_{SP}^{(1)}(\tau)$
are the degree of first-order coherence of Toptica laser and single photons, respectively.
Single photons inherit the coherence time ($>10\ \mu$s) from the cw Ti:saphhire laser (M SQUARED)
which has longer coherence time than that of the Toptica laser, so we have $|\mathrm{g}_{SP}^{(1)}(\tau)|=1$.
From the measured HOM dip width, we get the coherence time $\tau_L=150\ $ns for the Toptica laser, which
is also the mutual coherence time between Toptica laser and single photons.

A coincident event involves two photons detected by each of two detectors.
The first term in Eqs. (\ref{eq1}) and (\ref{eq2}) comes from the coincidence events for two
photons from the laser. The second term comes from the coincidence
events for two photons from single photons. The third term counts in the coincidence
events for two photons with one from the laser and one from single photons.
The fourth term in Eq. (\ref{eq2}) is due to the fourth-order interference between two photons
(one from laser and one from single photons), which essentially results from the
second-order interference between two photons within the mutual coherence time (see Supplementary Information).
Eqs. (\ref{eq1}) and (\ref{eq2}) fully reproduce the data in Figs. 2(d)-(e) and Fig. 3(a) [also see Fig. S1, Supplementary Information],
and yield the mode overlap $V_0=0.85$.

\begin{figure}[ht]
\centering
\includegraphics* [bb= 132 330 445 521, clip, width=8.5cm, height=5.2cm]{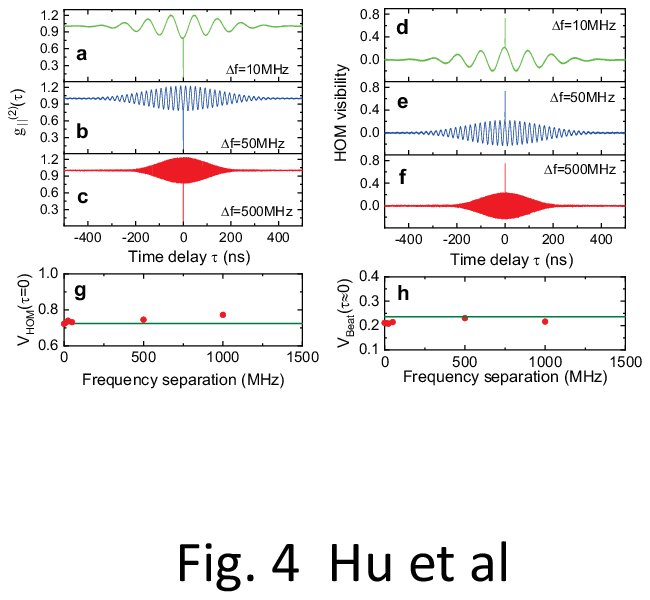}
\caption{(color online) Cross-correlation measurements in parallel-polarization configuration at different frequency separations which
are larger than the single-photon linwidth ($<100\ $kHz):
(a) $\Delta f=10\ $MHz; (b) $\Delta f=50\ $MHz; (c) $\Delta f=500\ $MHz. HOM visibility at different frequency separations:
(d) $\Delta f=10\ $MHz; (e) $\Delta f=50\ $MHz; (f) $\Delta f=500\ $MHz.
(g) $V_{HOM}(0)$ versus the frequency separation between laser light and single photons.
(h) The peak of the broad background visibility versus the frequency separation between laser light and single photons.
In both (g) and (h), red dots represent the experimental data and solid lines
represent the calculated results using Eq.(\ref{eq3}). }
\label{fig4}
\end{figure}

From Eqs. (\ref{eq1}) and (\ref{eq2}), we obtain the HOM visibility
\begin{equation}
V_{HOM}(\tau)=\frac{2\eta V_0e^{-|\tau|/\tau_L} \cos (2\pi\Delta f\tau)}{\eta^2+2\eta+\mathrm{g}_{SP}^{(2)}(\tau)}
\label{eq3}
\end{equation}
Eq. (\ref{eq3}) fully reproduces the data in Fig. 2(f) and Fig. 3(b) [also see Fig. S1, Supplementary Information].
Moreover, Eq. (\ref{eq3}) predicts $V_{HOM}(0)$ reaches its maximum
at $\eta=\sqrt{\mathrm{g}_{SP}^{(2)}(0)}=0.17$ where we take $\mathrm{g}_{SP}^{(2)}(0)=0.03$ and
the peak value of the broad background reaches its maximum at $\eta=1$, which
is verified by the experimental results in Fig. 3(b).

From the derivation of Eqs. (\ref{eq1}) and (\ref{eq2}) (see Supplementary Information) and the full
agreement between experiments and theory, we conclude that TPI as the fourth-order interference arises from the
second-order interference between two photons within the mutual coherence time.
Following this picture, TPI is not related to the photon indistinguishability and TPI should be
observable for distinguishable photons, which is verified by further experiments (see Fig. 4).
When the frequency separation between laser and single photons is several orders-of-magnitude (up to $10^4$)
larger than their linewidths ($<100\ $kHz), the TPI visibility $V_{HOM}(0)=72\%$ and the beat visibility $22\%$
[see Figs.4(b)-4(d)] remain the same as that for indistinguishable photons at $\Delta f=0$
[see Fig. 3(f)].

\begin{figure}[ht]
\centering
\includegraphics* [bb= 130 507 506 742, clip, width=8.5cm, height=5.3cm]{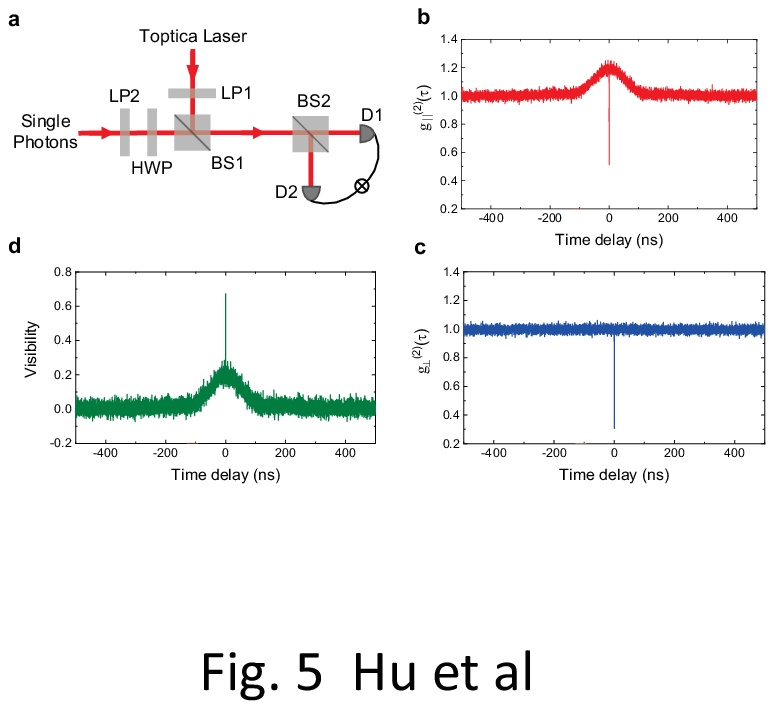}
\caption{(color online) Photon bunching induced by interference. (a)
Experimental setup to measure the bunching of two photons on BS1. LP1 and LP2: linear polarizers, HWP: half-wave plate,
BS1 and BS2: 50/50 non-polarizing beam splitters, D1 and D2: superconducting nanowire single photon detectors.
(b) Auto-correlation function $\mathrm{g}^{(2)}_{\parallel}(\tau)$ versus the time delay.
(c) Auto-correlation function $\mathrm{g}^{(2)}_{\perp}(\tau)$ versus the time delay .
(d) Visibility versus the time delay. The visibility is defined as $V(\tau)=[\mathrm{g}^{(2)}_{\parallel}(\tau)-\mathrm{g}^{(2)}_{\perp}(\tau)]/\mathrm{g}^{(2)}_{\perp}(\tau)$.
In (b)-(d), the photon frequency separation is kept at $\Delta f=0\ $MHz and the intensity ratio of laser
light to single photons is fixed to $0.2$.}
\label{fig5}
\end{figure}

Based on the above discussions, we see that the broad HOM dip induced by TPI occurs
on the time scale of the mutual coherence time ($150\ $ns) between laser and single photons,
different from the sharp single-photon antibunching dip on the time scale of photon correlation time ($115\ $ps).
In analogue to the HOM dip, the HOM bunching induced by interference
also occurs on the time scale of the mutual coherence time (see Fig. 5).
Corresponding to the sharp single-photon antibunching dip [see Figs. 5(b) and 5(c)],
a sharp visibility peak at $\tau=0$ is observed on the broad HOM bunching background
[see Fig. 5(d)]. Fig. 5(b)-(d) is fully reproduced by
Eqs. (S7) and (S8) in Supplementary Information and the calculated results are presented in Fig. S2.
These results reveal that the HOM bunching, i.e., the well-known photon bunching is rooted in
path indistinguishability, not photon indistinguishability.

To summarize, we have demonstrated the first observation of TPI between distinguishable
photons with their frequency separation up to $10^4$ times larger than their linewidths.
We coherently convert laser light into single photons with ultralong coherence
time ($>10\ \mu$s, five orders of magnitude longer than the photon correlation time)
inherited from the laser using a single QD coupling to an
optical microcavity. We perform time-resolved TPI between such coherent single photons
and an independent laser. We observe a maximum TPI visibility of $72\%\pm 2\%$
well above the $50\%$ classical limit and simultaneously
a broad visibility background and a classical beat visibility of less than $50\%$.
We find these visibilities are independent of the photon frequency separation and
show no difference between indistinguishable and distinguishable photons.
Using a general wave superposition model, we derive the cross-correlation functions
which fully reproduce and explain the experiments. Based on these results,
we uncover the essence of TPI that as the fourth-order interference TPI arises from the
second-order interference of two photons (indistinguishable or distinguishable)
arriving on the beam splitter within the mutual coherence time, and correspondingly
photon bunching takes root in path indistinguishability, not photon indistinguishability.

Our work provides new insights into TPI and has great implications in both quantum
optics and TPI-based photonic quantum technologies, such as long-distance TPI \cite{zhai22, you22},
asynchronous Bell-state measurement \cite{hu11, bhaskar20},
asynchronous measurement-device-independent quantum key distribution \cite{xie22, zeng22},
and boson sampling \cite{seron23, wang19}. Interfacing weak lasers commonly used in quantum
cryptography with single photons required for quantum state transfer, quantum repeaters
and quantum memories could be a realistic route towards quantum internet \cite{kimble08, wehner18, lu21}.

This work is supported by the Beijing Natural Science Foundation
under the grant IS23069. Z.C. Niu is grateful to the National Key Technology R$\&$D Program of China
under the grant 2018YFA0306101 for financial support.

\end{document}